\documentclass{elsart}
\journal{New Astronomy}

\usepackage{natbib}

\usepackage{graphicx}
\usepackage{epsfig}

\usepackage{amssymb}

\def\wisk#1{\ifmmode{#1}\else{$#1$}\fi}

\def\deg    {\wisk{^\circ}}
\def\ddeg   {\wisk{{\rlap.}^\circ}}

\begin{document}

\begin{frontmatter}

%

\title{ARCADE: Absolute Radiometer 
for Cosmology, Astrophysics, and Diffuse Emission
\thanksref{SMD}}
\thanks[SMD]{This work was supported by the suborbital program
of the NASA Science Mission Directorate
under RTOP 188-02-54-01.}

\author[gsfc]{A. Kogut}
\ead[url]{http://arcade.gsfc.nasa.gov/}
\author[gsfc,ssai]{D. Fixsen}
\author[gsfc,ssai]{S. Fixsen}
\author[jpl]{S. Levin}
\author[gsfc,ssai]{M. Limon}
\author[gsfc,ssai]{L. Lowe}
\author[gsfc,ssai]{P. Mirel}
\author[jpl]{M. Seiffert}
\author[ucsb]{J. Singal}
\author[ucsb]{P. Lubin}
\author[gsfc]{E. Wollack}
\address[gsfc]{NASA Goddard Space Flight Center, Greenbelt, MD 20771}
\address[ssai]{Science Systems and Applications, Inc.}
\address[jpl]{Jet Propulsion Laboratory, Pasadena CA}
\address[ucsb]{University of California at Santa Barbara}

\begin{abstract}
The Absolute Radiometer for Cosmology, Astrophysics,
and Diffuse Emission (ARCADE)
is a balloon-borne instrument
designed to measure the temperature of the cosmic microwave background
at centimeter wavelengths.
ARCADE searches for deviations from a blackbody spectrum
resulting from energy releases in the early universe.
Long-wavelength distortions in the CMB spectrum
are expected in all viable cosmological models.
Detecting these distortions or showing that they do not exist
is an important step for understanding the early universe.
We describe the ARCADE instrument design,
current status,
and future plans.
\end{abstract}


\end{frontmatter}

\section{Introduction}
The cosmic microwave background 
has proven to be an invaluable probe of physical conditions
in the early universe.
Temperature anisotropy in the CMB records the density 
and gravitational potential
at an epoch when the fluctuations 
leading to the present-day large scale structure
were still in the linear regime.
The CMB spectrum, by contrast,
provides information on the {\em energetics},
including the critical period after recombination
when the currently-observed large-scale structures first formed.
Precise measurements of the CMB spectrum 
provide important information on the early universe.

The frequency spectrum of the cosmic microwave background (CMB) 
carries a history of energy transfer 
between the evolving matter and radiation fields
in the early universe. 
Energetic events
(particle decay, star formation) 
heat the diffuse matter which then cools 
via interactions with the background radiation, 
distorting the radiation 
spectrum away from a blackbody. 
The amplitude and shape of the resulting 
distortion depend on the magnitude and redshift of the energy transfer
\cite{zeldovich/sunyaev:1969,
burigana/etal:1991,
burigana/etal:1995}.

\begin{figure}[t]
\centerline{
\psfig{file=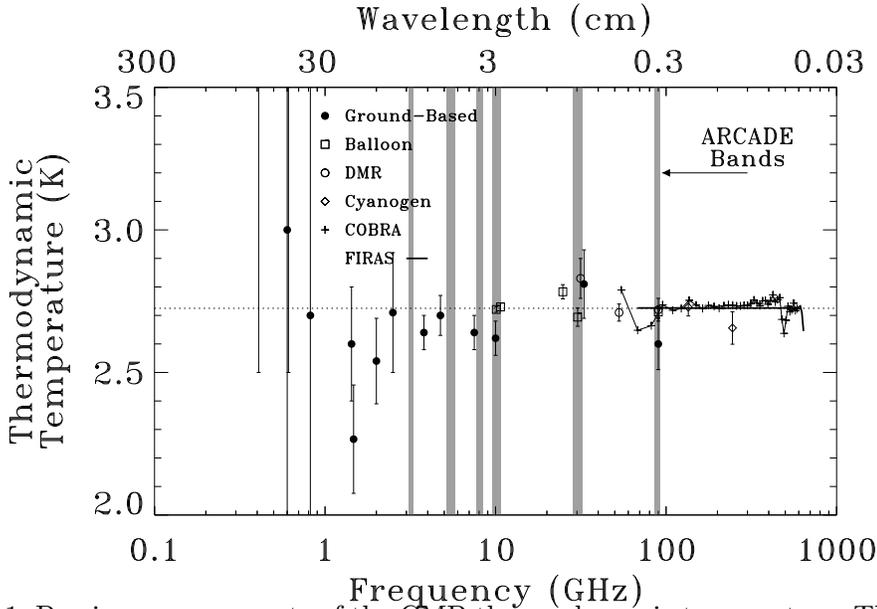,width=2.8in,angle=90}}
\caption{Precise measurements of the CMB thermodynamic temperature.
The dotted line represents a blackbody spectrum.
The gray bands show the ARCADE frequencies.
The CMB spectrum is poorly constrained at centimeter 
or longer wavelengths.}
\label{spectrum_fig}
\end{figure}

Measurements across the peak of the CMB spectrum
limit deviations from a blackbody
to less than 50 parts per million
\cite{mather/etal:1990,
gush/etal:1990,
fixsen/mather:2002}.
Direct observational limits at longer wavelengths, however, are weak:
distortions as large as 5\% could exist at wavelengths
of several centimeters or longer 
without violating existing observations
(Figure \ref{spectrum_fig}).
Plausible physical processes are expected to generate 
observable distortions without violating limits established
at shorter wavelengths.
Free-free emission from reionization
and subsequent structure formation
adds photons to the diffuse background,
creating a spectral distortion
$ \Delta T_{\rm ff} = T_{\gamma} Y_{\rm ff} / {x^2}$,
where $T_{\gamma}$ is the undistorted photon temperature,
$x$ is the dimensionless frequency $h \nu / k T_{\gamma}$,
\begin{equation}
Y_{\rm ff} = \int^z_0 ~\frac{ k[T_e(z) - T_{\gamma}(z)] }{ T_e(z) }
\frac{ 8 \pi e^6 h^2 n_e^2 g }
{ 3 m_e (kT_{\gamma})^3 \sqrt{6\pi m_e k T_e} }
\frac{dt}{dz^\prime} dz^\prime,
\label{Yff_definition}
\end{equation}
is the optical depth to free-free emission,
and g is the Gaunt factor 
\cite{bartlett/stebbins:1991}.
The distorted CMB spectrum is characterized 
by a quadratic rise in temperature at long wavelengths. 
The amplitude of the free-free signal 
depends on the column $\int n_e^2$ of ionized gas
and thus on the redshift $z_r$ 
at which the first collapsed objects formed.
Such a distortion {\it must} exist,
with predicted amplitude
of a few mK at frequency 3 GHz
\cite{haiman/loeb:1997,
gnedin/ostriker:1997,
oh:1999}.
Detection of the free-free distortion
would place important constraints
on the era of luminous object formation
and the extent of clumping in galactic halos.

The decay of massive particles
or other relics produced near the Big Bang
will also distort the CMB spectrum.
Energy released at an early epoch
to either charged particles or photons
will heat free electrons,
which then cool via Compton scattering from
the colder CMB photons.
For energy released at redshift $z < 10^4$
the gas is optically thin,
resulting in a uniform decrement
$ \Delta T_{\rm RJ} = T_{\gamma} (1 - 2y) $
in the Rayleigh-Jeans part of the spectrum 
where there are too few photons, 
and an exponential rise in temperature in the Wien region 
with too many photons. The magnitude of the distortion 
is related to the total energy transfer
$\Delta {\rm E} / {\rm E} = 4y$
where
\begin{equation}
y = \int^z_0 ~\frac{ k[T_e(z) - T_{\gamma}(z)] }{ m_e c^2} 
\sigma_T n_e(z) c \frac{dt}{dz^\prime} dz^\prime,
\label{compton_y_definition}
\end{equation}
is the dimensionless integral of the electron pressure 
$n_e k T_e$ along the line of sight,  
$m_e$, $n_e$ and $T_e$ are the electron mass, 
spatial density, and temperature,
$T_{\gamma}$ is the photon temperature, 
$k$ is Boltzmann's constant, 
$z$ is redshift,
and $\sigma_T$ denotes the Thomson cross section 
\cite{sunyaev/zeldovich:1970}.
Energy transfer at higher redshift $10^4 < z < 10^7$ 
approaches the equilibrium Bose-Einstein distribution, 
characterized by the dimensionless chemical potential 
$ \mu_0 = 1.4 \Delta {\rm E} / {\rm E}. $
Free-free emission thermalizes the spectrum at long 
wavelengths. Including this effect, 
the chemical potential becomes frequency-dependent,
\begin{equation}
\mu(x) = \mu_0 \exp(- \frac{2x_b}{x}),
\label{mu_vs_freq_eq}
\end{equation}
where $x_b$ is the transition frequency 
at which Compton scattering of photons 
to higher frequencies 
is balanced by free-free creation of new photons.
The resulting spectrum has a sharp drop 
in brightness temperature at centimeter wavelengths 
\cite{burigana/etal:1991}.

\begin{figure}[b]
\centerline{
\psfig{file=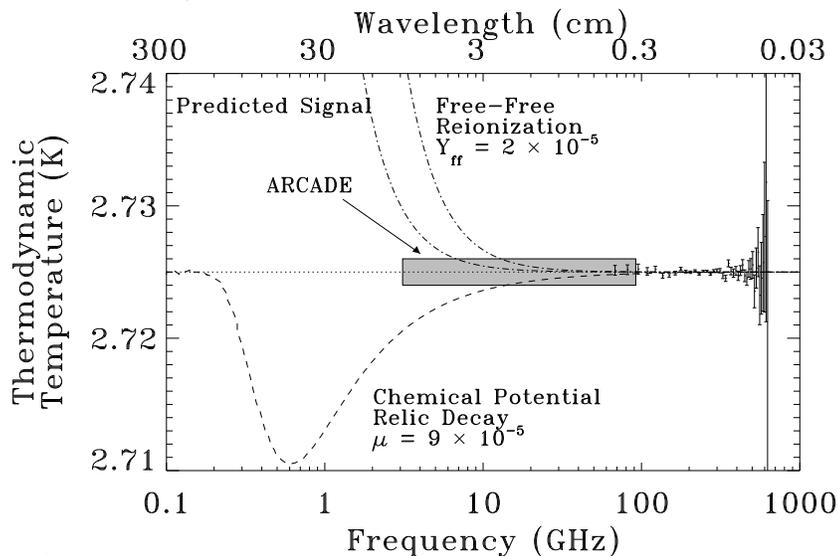,width=2.6in,angle=90}}
\caption{
Current 95\% confidence upper limits to distorted CMB spectra,
along with the predicted signal
from reionization and structure formation.
The FIRAS data and ARCADE 1 mK 
anticipated error box are also shown.
ARCADE has both the sensitivity 
and control of systematic errors
to measure the expected signal from
the high-redshift universe.}
\label{distortion_fig}
\end{figure}

A chemical potential distortion is a primary signature for
the decay of relics from GUT and Planck-era physics.
Such relics are expected to exist.
Growing evidence points to the existence of ``dark matter''
observed only through gravitational effects,
with density $\Omega_m = 0.22^{+0.01}_{-0.02}$
compared to the critical density
\cite{spergel/etal:2006}.
A leading candidate for dark matter composed of
weakly interacting massive particles
is the lightest supersymmetric particle $\chi$,
the neutralino
\cite{jungman/etal:1996}.
Neutralinos annihilate primarily into quark-antiquark pairs,
which in turn cascade to photons, electrons, positrons, and neutrinos. 
The annihilation rate 
$
\Gamma = n_\chi n_{\bar \chi} 
\langle \sigma v \rangle,
$
where $\langle \sigma v \rangle ~\sim 6 \times 10^{-26}$ 
cm$^3$ s$^{-1}$ at freeze out
and the neutralino space density
$n_\chi$ varies as $z^3$
from the overall expansion of the universe.
Gamma radiation 
from neutralino annihilation
in the Galactic dark matter halo
has been proposed to explain the excess flux
observed by the EGRET mission
\cite{fornengo/etal:2004,valle:2004}.
A chemical potential distortion to the CMB spectrum
is an inevitable result of neutralino annihilation:
annihilations sufficient to produce the EGRET excess
would necessarily produce a chemical potential distortion.
Measurements of the CMB spectrum
can thus provide important constraints
on the mass, cross section, and 
decay modes of dark matter candidates.

Figure \ref{distortion_fig}
shows current upper limits to
spectral distortions at long wavelengths.
The CMB spectrum is poorly constrained 
at centimeter or longer wavelengths,
where observable signals
from reionization
or relic decay
are expected to exist.
New measurements with mK accuracy
can provide important checks
for both cosmology and high-energy physics.
The Absolute Radiometer for Cosmology, Astrophysics,
and Diffuse Emission (ARCADE)
is a balloon-borne instrument 
to provide precisely such new data.

\section{Instrument Description}
\begin{figure}[b]
\centerline{
\psfig{file=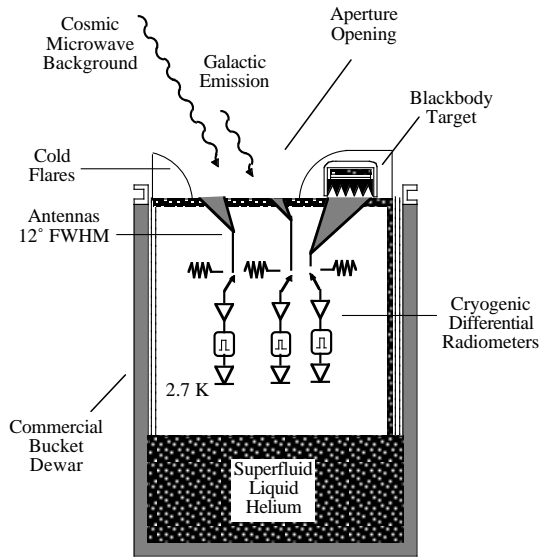,width=2.8in}}
\caption{
Schematic drawing of the ARCADE instrument.}
\label{inst_schematic}
\end{figure}

ARCADE is a balloon-borne,
fully cryogenic,
double-nulled instrument
to compare the temperature of the sky
to a precision on-board blackbody calibrator.
Figure \ref{inst_schematic} shows a schematic of the instrument.
It consists of 7\footnote{
Two independent radiometers
operate at 30 GHz,
with identical amplification chains
but different beam sizes
as a cross-check on emission from the reflector hiding the balloon
and flight train.}
narrow-band cryogenic radiometers
($\Delta \nu / \nu \sim 10\%$)
with central frequencies $\nu$ = 3.3, 5.5, 8.1, 10, 30, and 90 GHz
chosen to cover
the gap between full-sky surveys at radio frequencies ($\nu < 3$ GHz)
and the FIRAS millimeter and sub-mm measurements ($\nu > 60$ GHz).
Each radiometer measures the difference in power
between a beam-defining antenna (FWHM 11\ddeg6)
and a temperature-controlled internal reference load.
An independently controlled blackbody calibrator 
(emissivity $\epsilon > 0.9999$) is located on the aperture plane,
and moves to cover each antenna in turn,
so that each antenna alternately views the sky 
or a known blackbody.
The calibrator temperature can be adjusted 
to null the sky-antenna signal difference.
ARCADE thus measures small spectral shifts about a precise blackbody,
greatly reducing dependence on instrument calibration and stability.
The calibrator, antennas, and radiometer front-end amplifiers
are all maintained near thermal equilibrium with the CMB.
Boiloff helium vapor,
vented through the aperture,
forms a barrier between the instrument and the atmosphere,
allowing operation in full cryogenic mode
and greatly reducing thermal gradients within the instrument.

\begin{figure}[t]
\centerline{
\psfig{file=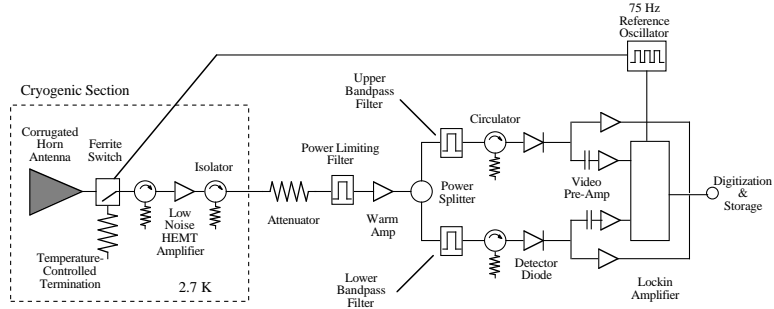,width=4.0in}}
\caption{
Block diagram of ARCADE radiometers.}
\label{block_diagram}
\end{figure}

Figure \ref{block_diagram} 
shows a block diagram of the ARCADE radiometers.
Each radiometer consists of a cryogenic front end
with a high electron mobility transistor (HEMT) 
direct-gain receiver,
switched at 75 Hz
for gain stability
between a wavelength-scaled corrugated conical horn antenna
and a temperature-controlled internal load.
We switch between the sky horn and internal load
using latching ferrite switches (8, 10, 30, and 90 GHz)
or coaxial MEMS switches (3 and 5 GHz)
maintained near the CMB temperature
to reduce effects of insertion loss to negligible levels.
The radiometer back ends are housed 
in a temperature-controlled module mounted 
outside the dewar,
with stainless steel waveguide (30 GHz and above) 
or coax (10 GHz and below)
providing the RF link between the cryogenic 
and room-temperature components.
The back end of each radiometer is split into 
two frequency sub-channels:
one covering the lower half of the front-end passband 
and an independent channel
covering the upper half of the front-end passband.
Video preamplifiers following each detector diode
separately amplify the dc 
and ac portions of the signal,
proportional to the
total power on the diode
and the antenna-load temperature difference, respectively.
A lockin amplifier demodulates the switched (ac) signal
and integrates for one second
to produce an output proportional to the difference in power
between the antenna and the internal load.
The lockin output and total power signals
are digitized at 1 Hz
and written to an on-board recorder
before being telemetered to the ground.

\subsection{Internal Loads}
The internal reference load has a dual purpose:
it provides a stable cold reference for the fast gain chop,
and can be adjusted in temperature during flight
to eliminate the radiometric offset 
resulting from imbalance 
within the switch or between the two arms of the radiometer.
An ideal load would have negligible reflection
and be isothermal so that its radiometric signal
could be computed from a single thermometer
connected to the absorbing element.
We come close to this ideal
using a thin layer of cryogenic-compatible microwave absorber
cast inside a copper waveguide.
The absorber consists of stainless steel powder
mixed with a commercially available epoxy
and can be tuned for the desired electromagnetic properties
\cite{wollack/etal:2006a}.
The absorber is thermally anchored 
within a high-thermal-conductivity waveguide,
which in turn in surrounded by foam insulation
with RF connection through a short stainless steel waveguide section.
A resistive heater mounted on a coaxial copper tube 
surrounding the foam layer
provides thermal control,
with cooling through a thin copper wire sunk to the 1.55 K 
liquid helium (LHe) bath.
Neither the heater nor the cooling wire are attached directly to the
copper waveguide, which acts as an open thermal circuit to minimize
thermal gradients across the absorber.
We measure power reflection -30 dB or better
across the band for each load
\cite{wollack/etal:2006b}.
Differential corrections for standing waves
are thus reduced below 1 mK for load setpoint changes 
as large as 1 K.

\begin{figure}[b]
\centerline{
\psfig{file=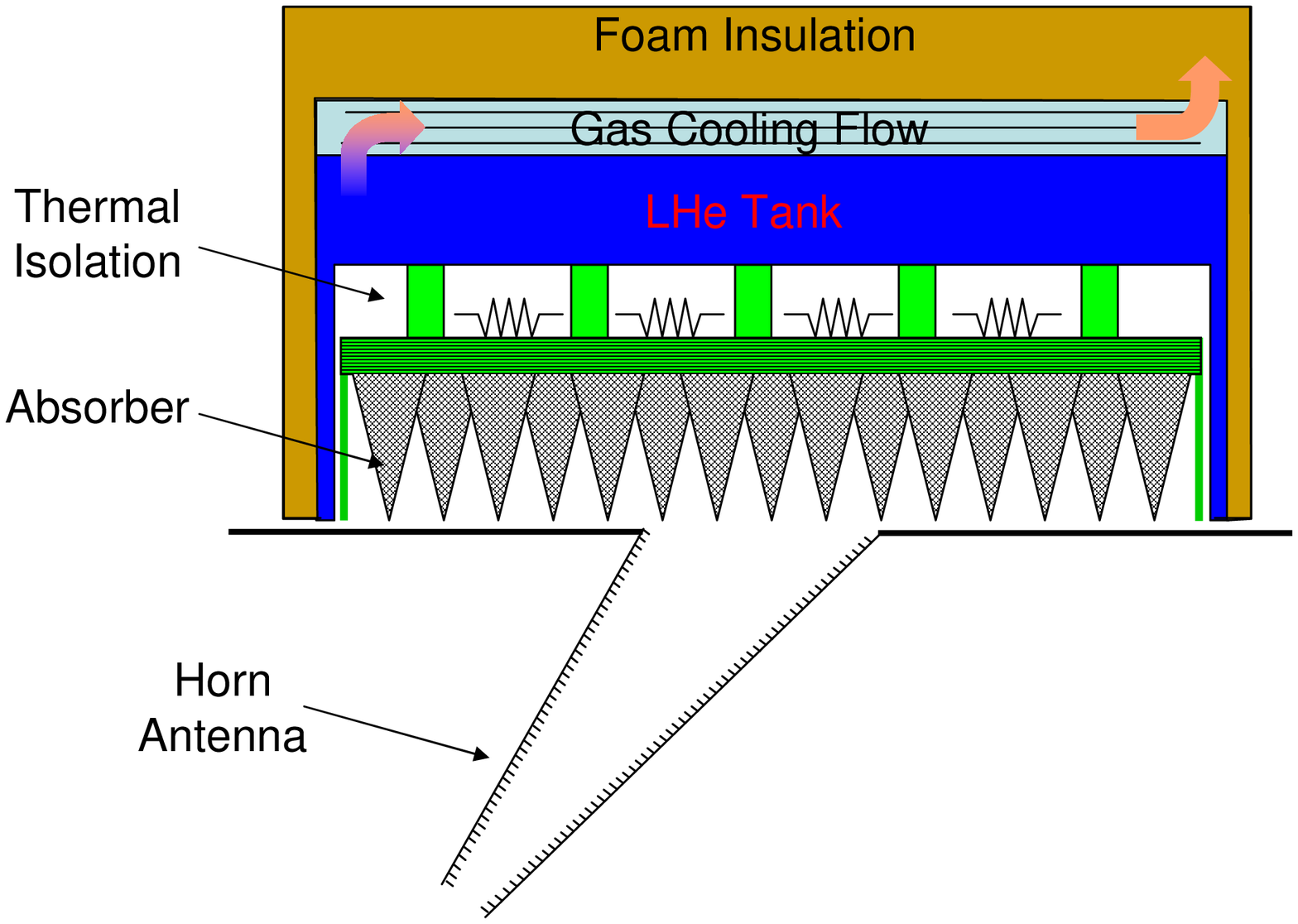,width=4.0in}
\psfig{file=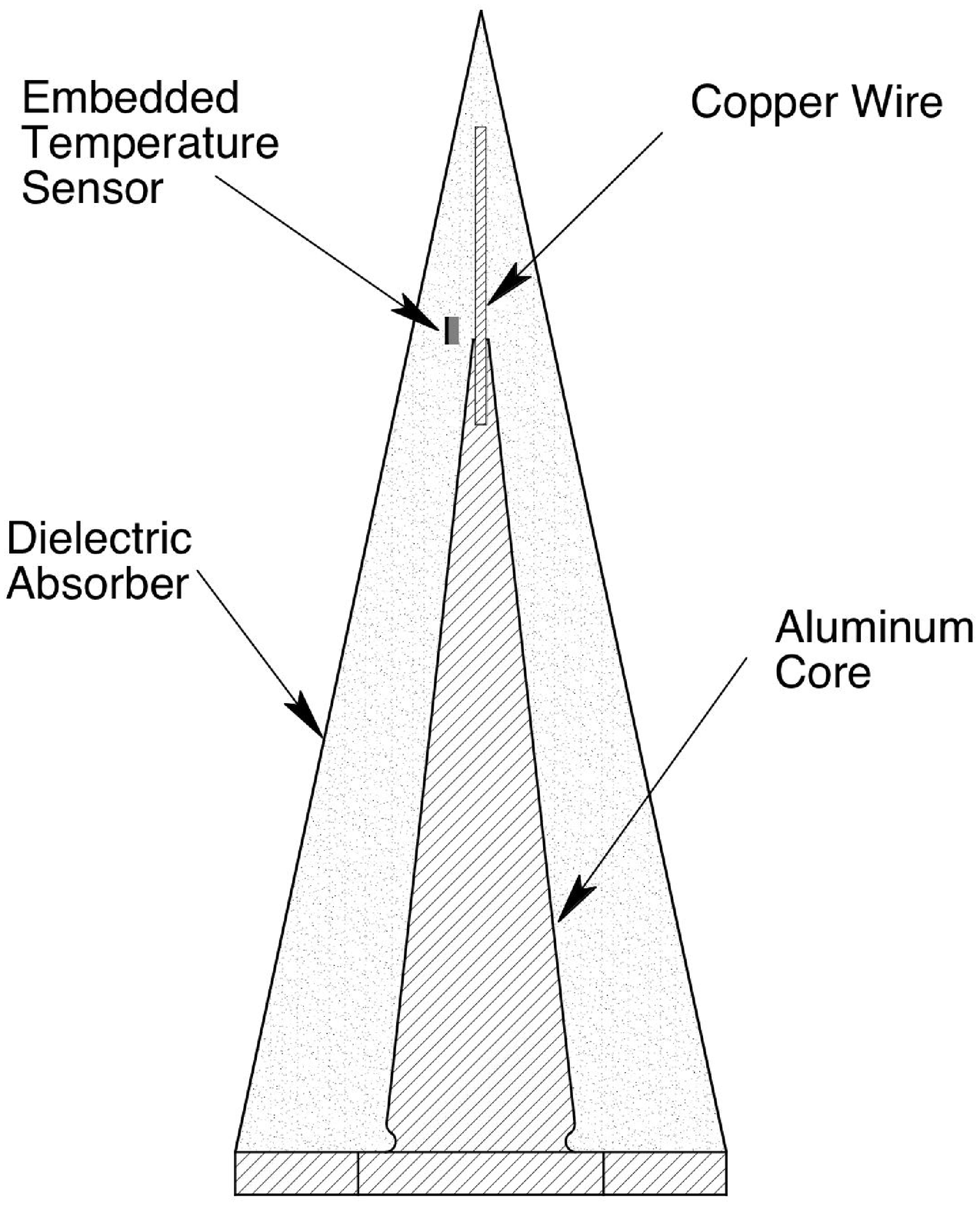,width=2.0in}}
\caption{Schematic of ARCADE external calibrator.
(left) An array of conical microwave absorbers  
is weakly coupled to a superfluid LHe reservoir.
Liquid-filled ``wings'' on the reservoir intercept
potential heat leaks from outside the dewar.
A metal/fiberglass thermal diffusion plate
serves to reduce the thermal ``footprint'' of heaters
at the absorber.
(right) A single absorbing cone
consists of a metal core
covered by microwave absorber.
Thermometers embedded within the absorber
monitor the temperature of the calibrator.
}
\label{target_fig}
\end{figure}

\subsection{External Calibrator}
The external calibrator provides an absolute radiometric calibration
by replacing the sky signal
with emission from a blackbody source at a known temperature.
Comparison of the sky to the external calibrator
provides a second level of nulling
in which fixed instrumental signals
from internal absorption or reflection 
will cancel.
The calibrator consists of a set of 298 absorbing cones
mounted on a thermal diffusion plate
and weakly coupled to a liquid helium reservoir
(Figure \ref{target_fig}).
Fountain-effect pumps lift superfluid liquid helium
from the main dewar to the calibrator;
thermal control is provided by heaters
mounted on the back side of the thermal diffusion plate.
The calibrator is mounted within a rotating carousel
above the cold aperture plane,
and can be moved to cover each antenna in turn.

The calibrator must satisfy several competing design requirements.
It must be black 
(emissivity $e > 0.999$ or
reflectivity $r = 1-e < 0.001$)
while maintaining a well-defined temperature.
It must completely fill the largest antenna aperture 
(620 mm diameter)
while remaining compact enough
to mount at the top of an open bucket dewar.
We meet these requirements using a composite design.
Each cone within the calibrator
is 88 mm tall and
consists of an aluminum core
coated with a 7 mm absorbing ``skin''
made from Steelcast absorber
\cite{wollack/etal:2006a}.
A copper wire extends from the inner aluminum cone
into the absorber tip.
Together, the aluminum and copper 
provide a high thermal conductivity path
to reduce thermal gradients within each cone,
allowing 95\% of the absorber volume to remain within 0.5 mK
of the base temperature
\cite{fixsen/etal:2006}.

The instrument operates in a null mode,
with the calibrator temperature close to the sky temperature
so that small reflections in the radiometer/calibrator system
cancel to first order.
Reflections from the calibrator create a systematic offset in the
sky-calibrator comparison
\begin{equation}
\Delta T_{\rm cal} = r (T_{\rm sky} - T_{\rm cal})
\end{equation}
proportional to the calibrator's power reflection coefficient $r$
and the temperature difference between the calibrator and the sky.
Requiring $\Delta T_{\rm cal} < 0.1$ mK
requires $r < 10^{-3}$.
Direct measurements of the antenna-calibrator combination
in flight configuration
show the power reflection r $<$ -42 dB
in the worst frequency band
\cite{fixsen/etal:2006}.
The calibrator completely covers the antenna aperture
and is mounted nearly flush with the aperture plane.
Measurements of the leakage through the 1 mm gap
between the calibrator and the antenna aperture
limit spillover past the calibrator to less than -50 dB.
Cold flares surrounding the aperture plane
redirect any residual spillover to blank areas of the sky.
Despite having an absorber less than one wavelength tall,
the ARCADE calibrator is demonstrated to be black 
to better than 0.01\%
across 5 octaves in frequency.

\subsection{Thermometry}
ARCADE's double-null design reduces the problem
of precise radiometric photometry
to one of simple thermometry:
we adjust the temperature of the internal load
to produce nearly zero output voltage on the
differential radiometer,
and then adjust the temperature of the external calibrator
until it matches the sky.
Provided the radiometer is linear
and the emissive components are nearly isothermal,
the sky temperature
can then be read off from the instrument thermometry.

We monitor cryogenic temperatures
using 4-wire ac resistance measurements
of 120 ruthenium oxide thermometers
read out every 1.067 seconds
\cite{fixsen/etal:2002}.
26 of the thermometers are embedded
in the absorbing skin of selected cones within the external calibrator.
An additional 9 thermometers measure temperatures
on the calibrator diffusion plate,
liquid helium tank,
and surrounding support structures.
21 additional thermometers
monitor 3 critical temperatures 
(antenna throat,
internal load,
and the Dicke switch/cryogenic HEMT amplifier)
on each of the 7 radiometers.
The remaining thermometers
monitor the temperature
of the aperture plane,
cold flares,
superfluid pumps,
liquid helium reservoirs
(main tank, external calibrator, aperture plane, 
and radiometer bases)
as well as the dewar walls
and the instrument support structure inside the dewar.

The in-flight thermometry 
compares the resistance of each thermometer
to a set of 4 calibration resistors
spanning the dynamic range of the thermometer resistances.
The calibration resistors are part of the readout electronics board,
located in a temperature-controlled enclosure.
One of the calibration resistors 
includes the electrical harness into the dewar
to monitor possible effects from electrical pickup
or stray capacitance.
The flight software uses a lookup table
to infer a temperature for each thermometer
using temperature-resistance curves
derived from ground tests.

\begin{figure}[b]
\centerline{
\psfig{file=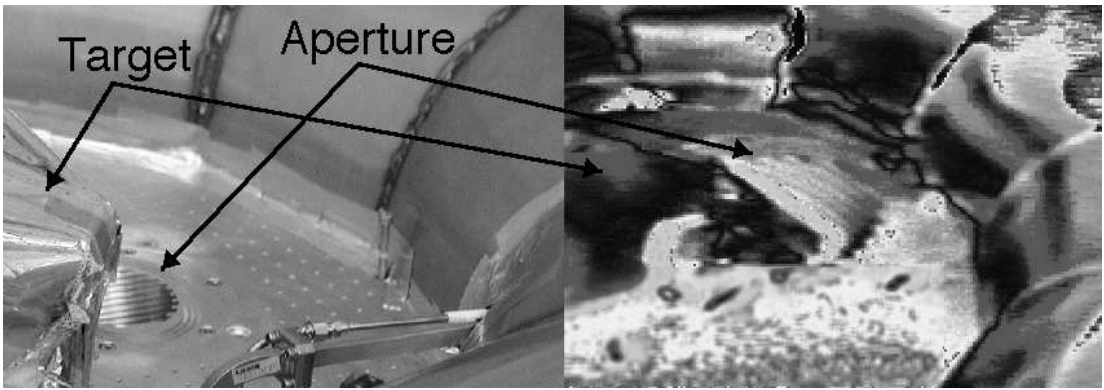,width=2.0in,angle=270}}
\caption{Aperture plane from 2001 flight 
with calibrator and antenna aperture.
(left) 30 GHz aperture during ground testing.  
The calibrator is visible in the left foreground;
the cold flares are in the background.
(right) Video still taken during 2001 flight,
30 minutes after opening ascent lid.
The corrugations of the 10 GHz aperture
are visible behind the calibrator in the foreground --
nitrogen condensation is not a major problem.}
\label{video_still}
\end{figure}

The instrument is mainly a transfer standard
to compare the sky
to the external calibrator;
precise knowledge of the absolute temperature
is required only for the external calibrator.
We build low-mass ruthenium oxide resistance thermometers
and establish the resistance to temperature calibration for
each thermometer in the external calibrator
by measuring the 
resistance of the embedded thermometers
while stepping the temperature of the instrumented cones
from 1.3 K to 15 K.
A separate NIST-calibrated resistance thermometer
of identical design
monitors the temperature throughout the calibration
to transfer the NIST calibration to the flight thermometers.
Since the thermometers are calibrated {\em in situ},
the small self-heating is inherently included.
This calibration has proved to be stable with respect to time.
Observations of the superfluid helium transition at 2.1768 K
are easlily observable in the calibration data
and provide a convenient cross-check on the 
resistance-temperature calibration.
The calibration is stable 
within 0.3 mK across 8 independent calibrations 
over 4 years,
providing both a cross-check on the absolute thermometry
and a limit to possible systematic drifts in the thermometer 
resistance-to-temperature curves
\cite{kogut/etal:2004a}.

\subsection{Antennas}
Each receiver is fed by a corrugated conical horn antenna
scaled in wavelength to produce identical 
11\ddeg6 ~full width at half maximum 
beam shape in each frequency channel
To avoid convective instabilities in the helium vapor barrier,
the instrument should remain vertical during observations.
We reconcile this requirement 
with the need for sky coverage 
(and avoiding direct view of the balloon)
by mounting the antennas at a 30\deg ~angle from the zenith,
slicing each antenna at the aperture plane.
The slice has minimal effect on the symmetry of the beams
\cite{singal/etal:2005}.

\subsection{Dewar and Cryogenic Aperture}
Active thermal control maintains the
external calibrator,
antennas,
Dicke switches,
and front-end amplifiers
at temperatures near 2.7 K,
in thermal equilibrium with the CMB signal.
To avoid correction for beam spillover onto the dewar walls,
we mount the antennas at the {\em top} of 
an open bucket dewar
1.5 m in diameter
and 1.8 m deep.
We further avoid corrections for emission or reflection
from a window over the cold optics
by using an open aperture
with no window between the cryogenic aperture
and the ambient environment.
Boiloff helium vapor,
vented through the aperture,
forms a barrier between the instrument
and the atmosphere
to allow operation in fully cryogenic mode.

Previous flights validate the open-aperture design.
The efflux of helium gas above the aperture
(roughly 5 m$^3$ s$^-1$ at 35 km float altitude)
does not completely eliminate condensation of atmospheric nitrogen
onto the cryogenic optics,
but does reduce it to levels compatible with 
the desired CMB observations.
Data from a 2001 flight of a 2-channel prototype payload 
show excess heat dissipation on the aperture plane,
consistent with an accumulation rate of approximately 
200 g hr$^{-1}$ of nitrogen onto the aperture,
a heat source of order 6 W
\cite{kogut/etal:2004b}.
Visual examination using an on-board video camera
confirms this slow accumulation of nitrogen ice,
with several mm of ``frost'' visible on the optics
an hour after opening the ascent lid
(Fig. \ref{video_still}).
Nitrogen ice is nearly transparent at cm wavelengths.
Since the ice cools to the same temperature as the rest of the optics,
modest accumulations affect only mechanical operations.
Analysis of the 2003 and 2005 flight data
shows no detectable radiometric signal
from ice accumulation
\cite{fixsen/etal:2004,singal/etal:2006}.

ARCADE has a large (kW) heater on the cold optics
to allow periodic de-icing of the external calibrator,
antenna aperture, and horn throat sections.
With the superfluid pumps off,
the aperture plane and calibrator have 
only weak thermal coupling to the dewar.
Raising the aperture plane and calibrator 
above 80 K for 15 minutes
suffices to remove accumulated nitrogen ice
with only a modest impact on the main liquid helium reservoir.
Atmospheric condensation is not a limiting problem.

\section{Instrument Status}
ARCADE was first selected for development in December 1999.
Flights of a 2-channel prototype in November 2001 and June 2003
demonstrated that a large (45 cm diameter) instrument
can be maintained at 2.7 K
at the {\em top} of a bucket dewar
without windows between the cold optics and the atmosphere
\cite{fixsen/etal:2004,
kogut/etal:2004b}.
Based on this experience,
a larger 6-channel instrument flew in July 2005.
This second-generation instrument
extended the frequency coverage to the longer wavelengths
needed to reach the main science goals,
while incorporating improvements in the cryogenic engineering
suggested by the first-generation instrument.
A gearbox mechanical failure during the 2005 flight
allowed only one antenna to view the sky,
sharply limiting the potential science return.
Engineering data showed that the remainder of the instrument,
including the radiometers and external calibrator,
performed well,
demonstrating the basic functionality 
of the second-generation instrument.

Science data from the 2005 flight
include several hours of sky data
with the 8 GHz radiometer,
operated in single-null condition
without the ability to view the external calibrator.
Even without the absolute reference,
the radiometer is sufficiently stable and linear
to allow a measurement of the CMB temperature
and Galactic foreground emission,
albeit at somewhat degraded accuracy
\cite{singal/etal:2006}.
We have since modified payload
to prevent a recurrence
and will re-fly the modified payload in July 2006.

The ARCADE design minimizes most sources of systematic error.
The leading uncertainty from the 2006 flight
is expected to result from small thermal gradients in the calibrator
as heat flows from the calibrator back plate
through the absorber
to the colder aperture plane below.
Future flights will reduce this source of uncertainty
by modifying the cryogenic aperture
to allow active thermal control of the
aperture plate and antenna mount structure.
The metal plate mounting the antenna apertures
is large (nearly 2 m$^2$ area) and exposed,
with complicated interactions through turbulent gas layers
to both the liquid helium below
and the ambient atmosphere above.
The large efflux of helium gas
precludes ground tests in a thermal vacuum chamber,
requiring actual flight data 
to allow engineering of incremental improvements.
A series of engineering tests during the 2006 flight
will provide thermal data
to guide modifications to the aperture plane
allowing active thermal control for a flight in 2007.


\end{document}